\documentclass[10pt, letterpaper, twocolumn]{article}

\usepackage{jstprove_whitepaper}

\title{JSTprove: Pioneering Verifiable AI for a Trustless Future}

\author{\vspace{1em}
  {\normalsize Jonathan Gold,} {\normalsize Tristan Freiberg,} {\normalsize Haruna Isah, and} {\normalsize Shirin Shahabi}\\
  {\normalsize [\href{mailto:jonathan@inferencelabs.com}{jonathan,} \href{mailto:tristan@inferencelabs.com}{tristan,} \href{mailto:haruna@inferencelabs.com}{haruna, and}  \href{mailto:shirin@inferencelabs.com}{shirin}]}@inferencelabs.com\\
  {\normalsize Inference Labs Inc.} 
}

\date{\today}

\begin{document}

\twocolumn[
\begin{@twocolumnfalse}
\maketitle
\begin{abstract}
\noindent
The integration of machine learning (ML) systems into critical industries such as healthcare, finance, and cybersecurity has transformed decision-making processes, but it also brings new challenges around trust, security, and accountability. 
As AI systems become more ubiquitous, ensuring the transparency and correctness of AI-driven decisions is crucial, especially when they have direct consequences on privacy, security, or fairness. 
Verifiable AI, powered by Zero-Knowledge Machine Learning (zkML), offers a robust solution to these challenges. 
zkML enables the verification of AI model inferences without exposing sensitive data, providing an essential layer of trust and privacy. 
However, traditional zkML systems typically require deep cryptographic expertise, placing them beyond the reach of most ML engineers. 
In this paper, we introduce JSTprove, a specialized zkML toolkit, built on Polyhedra Network's Expander backend, to enable AI developers and ML engineers to generate and verify proofs of AI inference. 
JSTprove provides an end-to-end verifiable AI inference pipeline that hides cryptographic complexity behind a simple command-line interface while exposing auditable artifacts for reproducibility. 
We present the design, innovations, and real-world use cases of JSTprove as well as our blueprints and tooling to encourage community review and extension. 
JSTprove therefore serves both as a usable zkML product for current engineering needs and as a reproducible foundation for future research and production deployments of verifiable AI.
\end{abstract}
\tableofcontents
\vspace{1em}

\end{@twocolumnfalse}
]

\section{Introduction}

As artificial intelligence (AI) and machine learning (ML) move from research labs into production systems, they now power social media feeds, product recommendations, medical decisions, and conversational tools \cite{chen2024zkml}. 
The ability to trust model outputs has become just as important as their accuracy. 
Within the AI industry, attention is shifting from model training to inference, the process of using trained models for making decisions or predictions on new data in real time. 
Industry analyses suggest that revenues from inference may soon rival or surpass those from training, reflecting its central role in real-world deployments. 
Inference now underpins decisions in domains such as healthcare diagnostics, fraud detection, cybersecurity monitoring, and financial forecasting, where even small errors or adversarial manipulations can carry high costs.

With the increasing complexity and influence of AI systems, the need for transparent and secure verification of AI-driven decisions has become paramount. 
On the one hand, clients are particularly vulnerable due to the sensitive nature of their data and are further concerned that ML service providers may employ under-performing models while presenting results that appear deceptively accurate. 
On the other hand, service providers face risks related to the theft or malicious compromise of their ML models \cite{peng2025survey}. 
Trust, while necessary, is no longer sufficient when decisions may impact privacy, security, or fairness. 
Verifiable AI, particularly zero-knowledge machine learning (zkML), offers a powerful solution that allows models to prove that inferences were computed faithfully according to the specified model without exposing sensitive data \cite{south2024verifiable}.

zkML uses zero-knowledge proofs (ZKPs) to verify that a model's outputs are derived by correct execution of the model's computation, without revealing private inputs. 
Growing concerns about data privacy have drawn increasing attention to ZKPs, leading to significant advances in both their theoretical foundations and practical implementations in ML \cite{sheybani2025zero}. 
However, current zkML methods often require specialized cryptographic expertise, making them difficult to use by developers accustomed to frameworks such as TensorFlow, PyTorch, Scikit-learn, and Keras. 
Designing ML operations with ZKP techniques is intricate and poses a barrier to adoption, as it requires bridging two distinct skill sets: machine learning and advanced cryptography.

In this paper, we present JSTprove, a zkML pipeline that orchestrates quantization, circuit construction, witness generation, and proof verification in an end-to-end workflow. 
Built on the Expander ecosystem \cite{expander2025}, it takes an ONNX model as input and produces a verifiable proof of inference, abstracting away protocol complexity behind a simple CLI. 
JSTprove demonstrates end-to-end verifiable inference on convolutional networks as an initial milestone, and its modular design is intended to extend naturally to a wider range of architectures. 
By integrating Expander's proving backend with our own quantization, circuit design, and pipeline automation, JSTprove provides developers with a usable framework for zkML. 
Looking ahead, JSTprove aims to address scalability, efficiency, and transparency challenges in AI verification, enabling eventual deployment in real-world applications. 
Through its open-source framework, JSTprove provides a simple CLI that reduces barriers to adoption while keeping workflows explicit and reproducible.

\section{Background and zkML Landscape}
\label{sec:background}

One of the most significant advancements in cryptography over the past decade has been the emergence of practical zero-knowledge proof (ZKP) schemes. 
A ZKP allows one party (the prover) to convince another (the verifier) that a computation was carried out correctly, without revealing the prover's private inputs or intermediate data. 
ZKPs are versatile across a wide range of protocols, supporting applications from authentication to blockchain smart contracts. 
Their security rests on established hardness assumptions from modern cryptography, while recent advances in protocol design have made proof generation and verification increasingly efficient. 
What sets ZKPs apart is their ability to certify the correctness of a computation without revealing any underlying inputs, a guarantee that distinguishes them from related privacy-preserving techniques such as homomorphic encryption and secure multiparty computation \cite{lavin2024survey}.

ZKPs are defined by three fundamental properties. 
Completeness means that if a statement about a computation is true, then an honest prover can always convince the verifier. 
Soundness ensures that if the statement is false, no cheating prover can convince the verifier otherwise, except with negligible probability. 
Zero-knowledge guarantees that, beyond learning whether the statement is true, the verifier gains no additional information about the prover's private inputs. 
Together, these properties enable verifiability of computations while preserving input privacy \cite{lavin2024survey, sheybani2025zero}. 
In complex applications such as verifiable machine learning inference, both prover and verifier may incur substantial computational costs. 
The prover typically represents the computation as an arithmetic circuit and then generates a proof that the circuit was evaluated correctly. 
Because these costs vary widely across protocols, understanding the trade-offs between different ZKP constructions is essential for practical deployment.

\subsection{ZKP Protocols}

Modern ZKP protocols differ in their degree of interactivity between prover and verifier, their setup requirements, and the trade-offs they make in efficiency. 
Thaler's work on verifiable computing \cite{thaler2022proofs} and the survey by Sheybani et al.\ \cite{sheybani2025zero} provide comprehensive taxonomies that classify protocols by their computational models and structural characteristics. 
Key attributes of a ZKP protocol include whether it is interactive or non-interactive, and the type of trusted setup required (circuit-specific, universal, or none). 
Other important factors are proof size, verification time, prover complexity, the underlying security assumptions (such as elliptic curve hardness, hash-function assumptions, or post-quantum candidates), scalability to large circuits or datasets, and whether the scheme is transparent (i.e., does not require a trusted setup).

Prominent families of protocols illustrate these trade-offs. 
Zero-Knowledge Succinct Non-Interactive Arguments of Knowledge (zk-SNARKs) produce very short proofs with fast verification, though most schemes require a trusted setup and rely on elliptic curve cryptography. 
Zero-Knowledge Scalable Transparent Arguments of Knowledge (zk-STARKs) avoid trusted setup altogether and achieve scalability using only collision-resistant hash functions, which also makes them plausibly post-quantum secure. 
MPC-in-the-Head (MPCitH) protocols simulate a secure multiparty computation internally: the prover generates multiple ``virtual'' parties' views, reveals a random subset, and thereby convinces the verifier of correct execution without disclosing private inputs. 
More recent VOLE-based constructions use vector-oblivious linear evaluation as a core building block to achieve efficient proof generation and verification.

The computational resources (time and memory) required to generate and verify proofs are critical factors in the practical deployment of ZKPs. 
Each construction presents distinct trade-offs in setup requirements, scalability, and security assumptions, making different protocols suitable for different privacy-preserving computation scenarios. 
Notable examples include Groth16 \cite{groth2018updatable}, Bulletproofs \cite{bunz2018bulletproofs}, SONIC \cite{maller2019sonic}, Libra \cite{xie2019libra}, Plonk \cite{gabizon2019plonk}, Halo \cite{bowe2019recursive}, Marlin \cite{chiesa2020marlin}, Spartan \cite{setty2020spartan}, and Hyrax \cite{patel2020lower}, each advancing the state of the art in succinctness, transparency, or efficiency.

Our contribution is not a new cryptographic protocol, but a developer-facing pipeline. 
We return to this after reviewing prior zkML frameworks.
For deeper technical discussions of ZKP protocols themselves, we refer readers to existing surveys \cite{sun2021survey,sheybani2025zero,thaler2022proofs,peng2025survey}.

\subsection{zkML Frameworks}

Real-world applications of zkML require translating high-level models into a representation suitable for proof systems. 
Neural network operations (e.g., convolutions, activations, pooling) must be expressed as algebraic constraints over a finite field, which are then compiled into arithmetic circuits. 
Constructing such circuits demands careful engineering, while efficient proof generation requires deep cryptographic expertise---barriers that limit adoption by typical ML developers. 
Libraries and pipelines mitigate this by abstracting circuit construction and proof generation. 
Most follow a two-step pattern \cite{thaler2022proofs}: a frontend compiles the model into a constraint system or arithmetic circuit, and a backend applies a ZKP protocol to verify its evaluation. 

Early work on zkML systems illustrates the same pattern. 
An example is \emph{ZEN}, introduced by Feng et al.\ \cite{feng2021zen}, which combined a quantization engine, circuit generators, and a scheme aggregator.
The system took pretrained PyTorch models, applied zk-SNARK-friendly quantization, and produced a quantized network to be proved using the Groth16 protocol \cite{groth2018updatable}. 
While ZEN demonstrated feasibility, it suffered from scalability limitations in prover runtime. 
Subsequent systems have explored alternative backends and integration strategies, seeking to balance developer accessibility with performance.

\emph{ZKML} \cite{chen2024zkml} introduced an optimizing compiler from TensorFlow to Halo2 circuits, demonstrating feasibility for realistic models such as convolutional vision networks and distilled GPT-2. 
Like other zkML frameworks, however, it faces challenges in supporting the full diversity of ML architectures and in managing the computational cost of circuit compilation for very large models or datasets.

An example of a lower-level approach is \emph{circomlib-ml} \cite{circomlibml2025}, a library of circuit templates for the Circom language. 
It provides gadgets for common ML components such as convolution, dense layers, pooling, batch normalization, and ReLU, along with polynomial activations designed to reduce constraint counts. 
The repository also includes MNIST case studies, illustrating how quantized neural networks can be expressed and tested directly in Circom. 
Although focused on building blocks rather than full pipelines, circomlib-ml demonstrates how ML primitives can be represented within general-purpose circuit DSLs.

\emph{EZKL} \cite{ezkl2025} emphasizes developer accessibility by providing an ONNX-to-Halo2 pipeline that abstracts most cryptographic details behind a command-line interface (CLI) and Python API. 
This allows practitioners to export models from PyTorch or TensorFlow into ONNX and generate a proof of correct inference without requiring deep knowledge of ZKPs. 
Built on the widely used Halo2 SNARK framework, EZKL benefits from existing tooling, audits, and support for on-chain verification, and has been applied in domains such as decentralized finance (DeFi) risk modeling \cite{ezkl2025-sentiment, ezkl2025-qiro}.

\emph{DeepProve} \cite{deeprove2025} is a zkML framework that supports multi-layer perceptrons (MLPs) and convolutional neural networks (CNNs) in the ONNX format. 
Its proving approach combines quantization with GKR-based subroutines such as sumchecks, lookup arguments, and polynomial accumulation. 
Each layer (e.g., dense, ReLU, maxpool) is expressed as a polynomial relation, and the prover demonstrates correct evaluation at randomly chosen points rather than proving every operation individually. 
Parameters are committed in advance, and intermediate claims are aggregated to reduce the number of polynomial openings required during verification. 
The system also includes requantization steps to ensure outputs remain within bounded ranges after multiplications. 

\emph{ZKTorch} \cite{chen2025zktorch} proposes a modular compiler that supports a wide range of neural architectures. 
Its design decomposes models into base cryptographic building blocks, enabling coverage of dozens of commonly used layers across CNNs, RNNs, and transformers. 
To improve efficiency, ZKTorch incorporates parallel proof aggregation techniques (via the Mira accumulation scheme \cite{beal2024mira}), which reduce proof size and speed up generation. 
Its transpiler and compiler jointly map high-level ML layers into these building blocks, aiming to provide scalable verification while preserving model privacy.

\emph{zkPyTorch} \cite{xie2025zkpytorch}, developed by Polyhedra Network, compiles PyTorch models into ZKP-compatible programs. 
It supports a variety of architectures, including CNNs, multi-layer perceptrons (MLPs), and transformers, and integrates with existing PyTorch and ONNX workflows. 
zkPyTorch is based on the GKR protocol and is powered by Polyhedra Network's Expander prover \cite{expander2025}, which combines GKR with polynomial commitment techniques to achieve scalability and performance.

The examples above are not exhaustive; many other frameworks and prototypes exist, but we focus here on a representative selection to illustrate the design space. 
For a comprehensive survey of zkML systems, we refer readers to Peng et al.\ \cite{peng2025survey}. 

Against this backdrop, JSTprove integrates Polyhedra Network's Expander ecosystem into an end-to-end ONNX-to-proof pipeline. 
Using the Expander Compiler Collection (ECC) \cite{expandercc2025} as a frontend for circuit construction and Expander \cite{expander2025} as the proving backend, JSTprove automates quantization, circuit construction, and artifact management, exposing them through a simple CLI (\texttt{compile}, \texttt{witness}, \texttt{prove}, \texttt{verify}). 
This design lowers the barrier for developers by hiding cryptographic details while keeping the workflow explicit and reproducible. 
At present, JSTprove supports a core set of neural network operations (GEMM, Conv2D, ReLU, MaxPool), but its modular architecture allows additional operators to be added over time.

\section{System Overview}
\label{sec:system}

JSTprove provides an end-to-end pipeline for turning an ONNX model into a provable inference. 
The process consists of the following stages:

\begin{enumerate}
    \item \textit{Model Import and Parsing.}
    Accepts ONNX models and parses their computational graphs, extracting layers, weights, and activations as preparation for circuitization.
    
    \item \textit{Quantization and Fixed-Point Conversion.}  
    Converts floating-point weights and activations into fixed-point integers using a simple scaling policy. This prepares model parameters for finite-field arithmetic in the ZK circuit while keeping behavior close to the original model.
    
    \item \textit{Circuit Compilation.}  
    Translates the quantized ONNX graph into an arithmetic circuit using the Expander Compiler Collection (ECC).
    Model operations are mapped layer by layer, with constraints inserted for inputs, outputs, and intermediate signals.
    
    \item \textit{Witness Generation.}  
    Given the circuit and quantized model, runs inference using the quantized model on user inputs and records the resulting outputs and auxiliary values as a witness. 

    \item \textit{Proof Generation.}  
    Uses Expander's GKR/sumcheck-based backend to generate a proof, with efficient verification, from the circuit and witness. 
    
    \item \textit{Verification.}  
    Checks the proof against all artifacts, ensuring that the claimed inference was performed faithfully.
    
    \item \textit{CLI and Output.}  
    The entire pipeline is exposed as a command-line tool. A single sequence of commands covers compilation, witness generation, proof production, and verification, making the workflow reproducible and transparent.
\end{enumerate}

\begin{figure}[ht]
\centering
\resizebox{\columnwidth}{!}{%
\begin{tikzpicture}[node distance=0mm, font=\small, every node/.style={font=\sffamily\small}]
\sffamily

\node[baseNode]                     (N1) at (0,0)                   {JSON input};           
\node[baseNode]                     (N2) at (\cstep,0)              {ONNX model};           

\node[baseNode]                     (N3) at (\cstep,-\rstep)        {Quantize model};           

\node[baseNode]                     (N4) at (\cstep,-2*\rstep)      {Translate NN graph $\to$ ZK constraints \\ (layer gadgets, I/O spec)};           

\node[baseNode]                     (N5) at (2*\cstep,-3*\rstep)    {Compile constraints $\to$ circuit.txt};           

\node[baseNode]                     (N6) at (0,-4*\rstep)           {Run quantized inference $\to$ outputs
Package inputs/outputs for witness gen};           

\node[baseNode]                     (N7) at (2*\cstep,-5*\rstep)    {Generate witness from packaged data $\to$ witness.bin};           

\node[baseNode]                     (N8) at (2*\cstep,-6*\rstep)    {Prove: circuit + witness $\to$ proof.bin};           

\node[baseNode]                     (N9) at (2*\cstep,-7*\rstep)    {Verify: circuit + proof};           

\draw[flow] (N1.south) -- (N6.north);
\draw[flow] (N2.south) -- (N3.north);
\draw[flow] (N3.south) -- (N4.north);
\draw[flow] (N7.south) -- (N8.north);
\draw[flow] (N8.south) -- (N9.north);

\draw[flow] (N4.south) to[out=-90, in=90, looseness=0.5] (N5.north);

\draw[flow] (N5.south) to[out=-90, in=0, looseness=0.9] (N6.east);

\draw[flow] (N6.south) to[out=-90, in=90, looseness=0.4] (N7.north);

\begin{scope}[on background layer]
  \node[groupBox, fit=(N1)(N2), inner sep=10pt] (G1) {};

  \node[groupBox, fit=(N3)(N4)(N6), inner sep=10pt] (G2) {};

  \node[groupBox, fit=(N5)(N7), inner sep=10pt] (G3) {};

  \node[groupBox, fit=(N8)(N9), inner sep=10pt] (G4) {};
\end{scope}
\node[font=\bfseries\footnotesize, align=center,
      fill=groupbg, draw=groupborder,
      inner sep=2pt, rounded corners=2pt]
   at (G1.center) {User};
\node[font=\bfseries\footnotesize, align=center,
      fill=groupbg, draw=groupborder,
      inner sep=2pt, rounded corners=2pt]
   at (G2.center) {JSTprove (orchestrator)};
\node[font=\bfseries\footnotesize, align=center,
      fill=groupbg, draw=groupborder,
      inner sep=2pt, rounded corners=2pt]
   at (G3.center) {ECC (circuits)};
\node[font=\bfseries\footnotesize, align=center,
      fill=groupbg, draw=groupborder,
      inner sep=2pt, rounded corners=2pt]
   at (G4.center) {Expander (prover)};
\end{tikzpicture}
}
\caption{JSTprove pipeline}
\label{fig:jstprove_pipeline}
\end{figure}
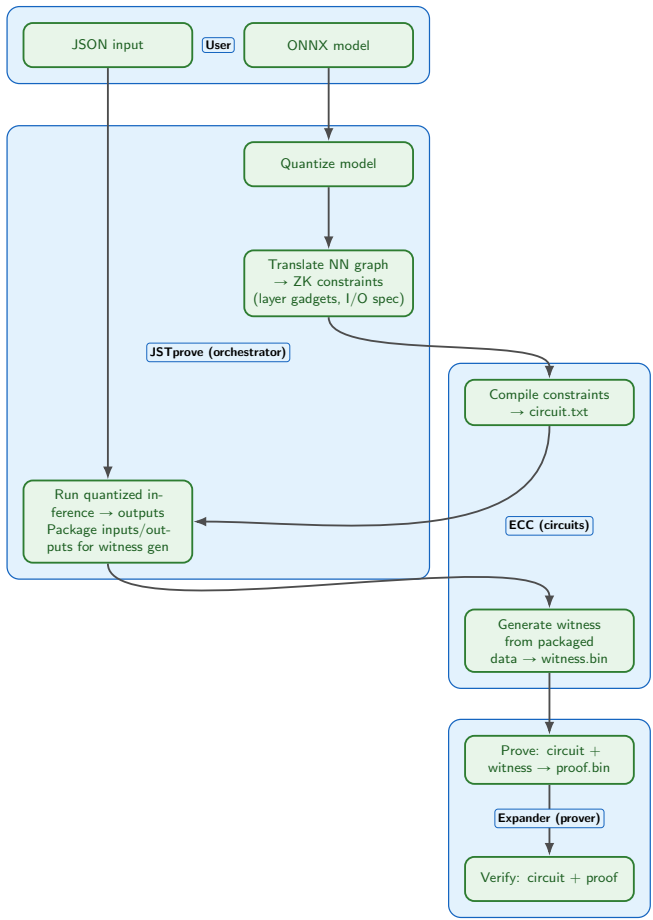 

At present, the system supports a core set of neural network layers and activations, including fully connected layers, 2D convolution (Conv2D), 2D max pooling (MaxPool2D), and ReLU.
The interface exposes four main commands:
{\small 
\begin{verbatim}
compile   # import, quantize, compile circuit
witness   # run quantized model, produce witness
prove     # generate proof
verify    # verify proof
\end{verbatim}
}
These stages may be invoked sequentially to run an end-to-end provable inference on any supported ONNX model. 
See Listing~\ref{lst:cli_demo} in the Appendix for a detailed example using the LeNet-style model.

\section{Circuit Design}
\label{sec:circuit-design}

We maintain an open \emph{zkML Blueprints} repository \cite{zkmlblueprints2025} containing mathematical proofs, pseudocode, and code using the Expander Compiler Collection (ECC) \cite{expandercc2025} Rust API for a core set of arithmetic circuits, including matrix multiplication, convolution, max pooling, and ReLU. 
The aim is not only to provide working implementations, but also to document the reasoning behind them in a form that is transparent, auditable, and open to community engagement. 
By publishing evolving specifications and correctness arguments, we enable external review and collaborative improvement, which increases confidence that the circuits behave as intended and lowers the barrier for others to build upon them.

As a representative primitive, we outline our design for \emph{range checking} via bitstring decomposition (to be optimized with lookup tables in future versions). 
The ability to prove that an integer lies within a prescribed interval serves as a fundamental building block for higher-level circuit gadgets: it enables ReLU activations (by exposing the sign of a value), max pooling (by enforcing comparisons), and quantized matrix multiplication (by checking the bounded remainder in fixed-point rescaling).

Suppose we wish to constrain an integer $x$ to the range $[-2^{\kappa-1},\,2^{\kappa-1}-1]$ using an arithmetic circuit over a prime field $\mathbb{Z}/p\mathbb{Z}$ with $2^{\kappa} \le p$. In the circuit, $x$ is represented by its least nonnegative residue $\bar{x} \in \{0,\ldots,p - 1\}$. 
Shifting to $x^{\sharp} = \bar{x} + 2^{\kappa - 1}$ places valid values in $[0,2^{\kappa} - 1]$. Conversely, if $x^{\sharp}$ lies in this interval, then $x$ is in the desired signed range, provided we adopt the usual convention that each integer is decoded from its balanced residue representative in $[-(p - 1)/2, (p - 1)/2]$.

To obtain a bitstring representation of $x^{\sharp}$, we use the unconstrained bitwise operations provided by ECC's Rust \texttt{RootAPI}. Repeated calls to \texttt{unconstrained\_bit\_and} and \texttt{unconstrained\_shift\_r} extract the $\kappa$ least significant bits $d_0,\ldots,d_{\kappa-1}$ of $x^{\sharp}$ in little-endian order.
We then enforce booleanity constraints
\[
d_i(d_i - 1) \equiv 0 \bmod p
\]
for each $i$, and reconstruct $x^{\sharp}$ by asserting
\[
x^{\sharp} \equiv d_0 + 2d_1 + \cdots + 2^{\kappa - 1} d_{\kappa - 1} \bmod p.
\]
The unconstrained bitwise extraction is efficient but not sound on its own; soundness is guaranteed by the combination of booleanity and reconstruction constraints, which bind the $d_i$'s to the unique binary expansion of $x^{\sharp}$ under the balanced-residue assumption.

Constraining an integer $x$ to a nonnegative interval such as $[0, 2^{\kappa} - 1]$ is a direct application of the range-check primitive and does not require the initial translation step used for signed ranges. In effect, we certify $x \ge 0$ by proving that $x$ lies in a dyadic interval anchored at $0$, with the right endpoint chosen large enough to accommodate all valid inputs. 
The same idea underlies comparisons: to enforce $x = \max\{a, b\}$ we require both $x - a \ge 0$ and $x - b \ge 0$, together with the selection constraint 
\[
(x - a)(x - b) \equiv 0 \bmod p,
\]
which forces $x$ to equal one of its arguments. The rectified linear unit $\mathrm{ReLU}(c) = \max\{c, 0\}$ is simply the special case with $b = 0$.

These simple patterns generalize further to max-pooling layers and to quotient-remainder checks in quantized arithmetic.
Together, these primitives form the foundation of the circuits used throughout JSTprove. 
While not yet optimized, they provide a clear baseline against which future refinements (such as lookup-based range checks) can be measured in practice.

\section{Quantization Strategy}
\label{sec:quantization}

To represent real-valued models in arithmetic circuits, we quantize all floating-point weights, activations, and inputs into integers. The guiding principle is to keep the arithmetic cheap (low-degree constraints in a prime field) while maintaining fidelity to the original floating-point computation.

We adopt a simple \emph{fixed-point} scheme. For a scaling factor $\alpha = 2^s$, each floating-point value $z$ is mapped to the integer
\[
\hat{z} = \lfloor \alpha z \rfloor.
\]
For negative $z$, the floor function means $\hat{z}$ may undershoot slightly, but this choice makes the mapping consistent across positive and negative domains. 
All quantized tensors are thus integer arrays of scale $\alpha$.

When two quantized values $a = \lfloor \alpha x \rfloor$ and $b = \lfloor \alpha y \rfloor$ are multiplied, their product $ab$ carries scale $\alpha^2$. To bring the result back to scale $\alpha$, we divide by $\alpha$:
\[
c = \frac{ab}{\alpha}.
\]
Since integer division discards the remainder, we enforce in the circuit that
\[
ab = \alpha q + r, \quad 0 \le r \le \alpha - 1,
\]
where $q$ is the rescaled output (our proxy for $\lfloor \alpha xy \rfloor$) and $r$ is the discarded remainder. The remainder bound is certified by a range check, reusing the bit-decomposition primitive from Section~\ref{sec:circuit-design}. In this way, quantization is reduced to a quotient-remainder check plus a range constraint.

A slight complication arises if $ab$ is negative, for we can only work with least nonnegative residues within the circuit. 
The standard workaround is a translation trick: assuming all valid products $ab$ lie within $[-\alpha 2^{\nu - 1},\alpha 2^{\nu - 1} - 1]$, we form
\[
ab + \alpha 2^{\nu - 1} = \alpha q^{\sharp} + r, \quad 0 \le r \le \alpha - 1,
\]
using ECC's \texttt{unconstrained\_int\_div} and \texttt{unconstrained\_int\_mod}.
The witnesses are a nonnegative quotient $q^{\sharp} \in [0,2^{\nu} - 1]$ and 
a remainder $r$. 
The circuit enforces
\[
ab \equiv \alpha q + r \bmod p, \quad q = q^{\sharp} - 2^{\nu-1},
\]
together with range checks for both $r$ and $q^{\sharp}$. These constraints ensure that $q$ indeed plays the role of $\lfloor \alpha xy \rfloor$ under the stated validity assumptions. 
Care must be taken in choosing $\nu$ and $\alpha$ so that wraparound modulo $p$ 
is avoided; see \cite{zkmlblueprints2025} for further details.

For simplicity, we currently restrict $\alpha$ to be a power of two. This ensures that multiplications and rescalings can be implemented with cheap shifts, and that range bounds align with dyadic intervals. 
More sophisticated quantization schemes (e.g., per-channel scales or lookup-based rounding) are possible and are left as future work.

At present, matrix multiplication is realized by directly verifying all of the quantized products in each sum-product expansion. 
Each individual product uses the quotient-remainder construction described above, with range checks enforced on both the remainder $r$ and the shifted quotient $q^{\sharp}$. 
This design is sound but constraint-heavy, since every scalar multiplication 
requires a bit-decomposition range check.

In future iterations we plan to integrate \emph{Freivalds' algorithm} \cite{Freivalds1977}, a 
classical randomized verification method for matrix products, into the circuit. 
Freivalds' trick reduces the verification of a full matrix multiplication to a small number of random linear checks, which can lower constraint costs. 

Finally, note that the bitstring decomposition required for the range check of $q^{\sharp}$ can be reused when a matrix multiplication is followed by a ReLU  activation. 
In this case, the most significant bit of the decomposition directly reveals the sign of the intermediate value, allowing the ReLU to be enforced without an additional decomposition. 
This simple fusion of \emph{MatMul+ReLU} is an important optimization for keeping circuit sizes manageable.

This quantization scheme is simple and sound but introduces significant overhead, making it a natural focal point for future optimization. 
The benchmarking results that follow should therefore be viewed as a baseline, establishing ground truth against which more efficient and probabilistic verification strategies can later be compared. 

\section{Benchmarking}
\label{sec:benchmarking}

In order to evaluate JSTprove's performance, we carried out a series of controlled benchmarks on two families of convolutional neural networks. 
Runtime measurements were recorded directly during benchmarking, while memory usage was measured externally at the operating system level using \texttt{psutil}, which reports peak resident set size (RSS), i.e., the maximum physical memory footprint observed during execution. 

All benchmarks were executed on a MacBook Pro (model FRW33LL/A, identifier Mac15,11) equipped with an Apple M3 Max chip, featuring 14 CPU cores (10 performance and 4 efficiency), 36 GB of unified memory, and running macOS Sonoma 14.7.2. 
Storage was provided by the built-in Macintosh HD.

\subsection{Depth Sweep}
\label{sec:depth_sweep}

Our first CNN family was generated via a ``depth sweep''. 
Table~\ref{tab:depth_time_mem} presents average time (in seconds) and memory (in megabytes) across three independent runs for each stage of the proving pipeline (circuit compilation, witness generation, proof generation, and verification) versus the number of model parameters. 

\begin{figure}[ht]
\centering
\begin{subfigure}[h]{0.48\columnwidth}
\centering
\includegraphics[height=0.5\textheight]{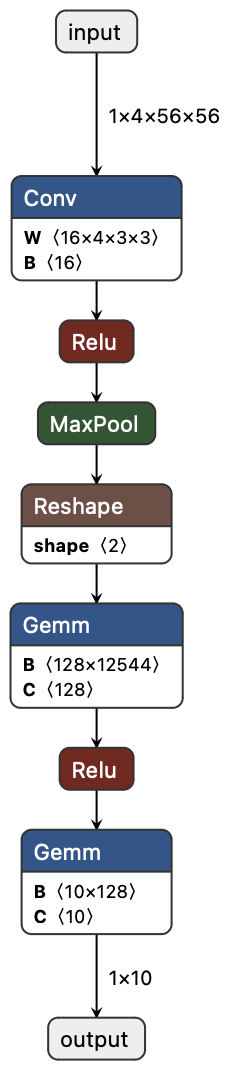}
\caption{Depth = 1}
\end{subfigure}%
\hfill
\begin{subfigure}[ht]{0.48\columnwidth}
\centering
\includegraphics[height=0.5\textheight]{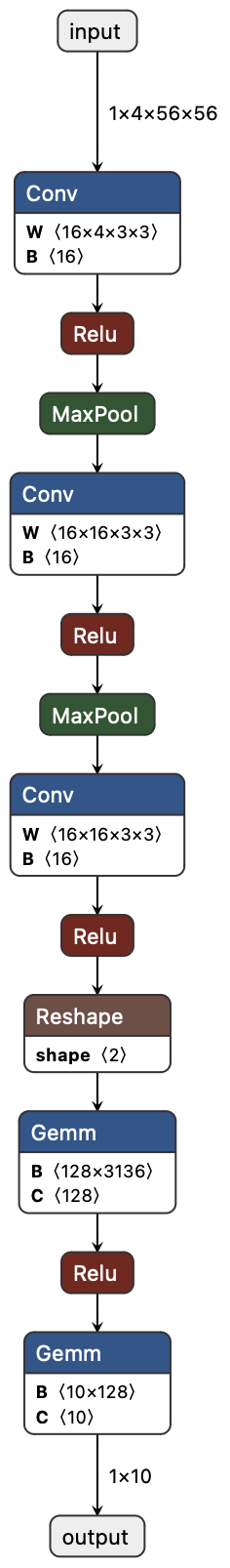}
\caption{Depth = 3}
\end{subfigure}
\caption{CNN structure for depth sweep}
\label{fig:cnn_diagram}
\end{figure}

Sixteen distinct models were benchmarked. 
These models follow the same design principles as LeNet, but depth is varied systematically: the first two blocks consist of \texttt{conv}--\texttt{ReLU}--\texttt{maxpool}, and any additional depth beyond this point is added by stacking further convolutional layers without additional pooling. 
Each network concludes with a reshape into a classifier tail of fully connected layers. 
Figure~\ref{fig:cnn_diagram} provides a visual representation of this architecture at depths 1 and 3.

\begin{table}[ht]
\centering
\textbf{Runtime (s) by phase}
\vspace{2pt}
\resizebox{\columnwidth}{!}{%
\begin{tabular}{rrrrrrrrrr}
\toprule
d & c & p & f & r & parameters & compile & witness & prove & verify \\
\midrule
1 & 1 & 1 & 1 & 2 & 1,607,642 & 179.93 & 5.79 & 13.90 & 8.80 \\
2 & 2 & 2 & 1 & 3 & 405,738 & 232.75 & 6.81 & 14.72 & 9.30 \\
3 & 3 & 2 & 1 & 4 & 408,058 & 246.89 & 7.05 & 14.89 & 9.52 \\
4 & 4 & 2 & 1 & 5 & 410,378 & 260.83 & 7.26 & 15.12 & 9.68 \\
5 & 5 & 2 & 1 & 6 & 412,698 & 278.47 & 7.50 & 15.24 & 9.79 \\
6 & 6 & 2 & 1 & 7 & 415,018 & 297.28 & 7.72 & 15.29 & 9.94 \\
7 & 7 & 2 & 1 & 8 & 417,338 & 313.35 & 8.12 & 15.49 & 10.01 \\
8 & 8 & 2 & 1 & 9 & 419,658 & 334.63 & 8.32 & 15.70 & 10.15 \\
9 & 9 & 2 & 1 & 10 & 421,978 & 353.59 & 8.60 & 15.80 & 10.26 \\
10 & 10 & 2 & 1 & 11 & 424,298 & 1075.20 & 9.19 & 15.86 & 10.37 \\
11 & 11 & 2 & 1 & 12 & 426,618 & 415.75 & 59.56 & 15.89 & 10.48 \\
12 & 12 & 2 & 1 & 13 & 428,938 & 408.25 & 9.41 & 16.31 & 10.65 \\
13 & 13 & 2 & 1 & 14 & 431,258 & 431.96 & 9.73 & 16.37 & 10.68 \\
14 & 14 & 2 & 1 & 15 & 433,578 & 459.44 & 10.00 & 16.57 & 10.92 \\
15 & 15 & 2 & 1 & 16 & 435,898 & 584.26 & 10.29 & 16.55 & 11.01 \\
16 & 16 & 2 & 1 & 17 & 438,218 & 603.39 & 10.51 & 16.77 & 11.08 \\
\bottomrule
\end{tabular}
}
\medskip\par
\textbf{Peak memory (GB) by phase}
\vspace{2pt}
\resizebox{\columnwidth}{!}{%
\begin{tabular}{rrrrrrrrrr}
\toprule
d & c & p & f & r & parameters & compile & witness & prove & verify \\
\midrule
1 & 1 & 1 & 1 & 2 & 1,607,642 & 23.194 & 4.452 & 6.700 & 4.871 \\
2 & 2 & 2 & 1 & 3 & 405,738 & 24.165 & 5.142 & 7.013 & 5.198 \\
3 & 3 & 2 & 1 & 4 & 408,058 & 24.408 & 5.224 & 7.166 & 5.375 \\
4 & 4 & 2 & 1 & 5 & 410,378 & 24.376 & 5.367 & 7.173 & 5.312 \\
5 & 5 & 2 & 1 & 6 & 412,698 & 24.614 & 5.640 & 7.344 & 5.550 \\
6 & 6 & 2 & 1 & 7 & 415,018 & 25.146 & 5.838 & 7.309 & 5.418 \\
7 & 7 & 2 & 1 & 8 & 417,338 & 24.943 & 5.931 & 7.433 & 5.349 \\
8 & 8 & 2 & 1 & 9 & 419,658 & 25.442 & 6.237 & 7.658 & 5.732 \\
9 & 9 & 2 & 1 & 10 & 421,978 & 25.371 & 6.640 & 7.701 & 5.711 \\
10 & 10 & 2 & 1 & 11 & 424,298 & 26.131 & 6.613 & 7.546 & 5.804 \\
11 & 11 & 2 & 1 & 12 & 426,618 & 25.877 & 6.638 & 7.921 & 5.927 \\
12 & 12 & 2 & 1 & 13 & 428,938 & 26.798 & 6.943 & 7.714 & 6.089 \\
13 & 13 & 2 & 1 & 14 & 431,258 & 26.728 & 7.179 & 7.890 & 6.064 \\
14 & 14 & 2 & 1 & 15 & 433,578 & 26.387 & 7.158 & 7.989 & 6.381 \\
15 & 15 & 2 & 1 & 16 & 435,898 & 26.552 & 7.272 & 8.010 & 6.396 \\
16 & 16 & 2 & 1 & 17 & 438,218 & 26.456 & 7.447 & 8.220 & 6.602 \\
\bottomrule
\end{tabular}
}
\caption{Depth sweep results: runtime and peak memory. Input size fixed at $h = w = 56$. The first two blocks are \texttt{conv}--\texttt{ReLU}--\texttt{maxpool}; additional depth adds convolutions without further pooling. $d$ is depth (number of conv blocks), and $c$, $p$, $f$, $r$ denote convolution, max-pool, fully-connected, and ReLU counts, respectively. Memory is peak RSS via \texttt{psutil}. \textit{parameters} is the ONNX parameter count.}
\label{tab:depth_time_mem}
\end{table}

Note that the depth-1 model has a substantially larger FC layer due to less pooling, making its parameter count atypically high.
With a single max-pooling layer and inputs of size $1 \times 4 \times 56 \times 56$, the model has approximately 1.6 million parameters. 
At depth 2, however, two rounds of pooling downsample the activations, leading to a much smaller fully connected input size, and thus only about 0.4 million parameters. 
Despite this reduction in parameter count, time and memory usage grow, because convolutional layers are expensive to circuitize: every additional convolution introduces thousands of multiplication and addition gates, which outweigh the reduction in fully connected weights. 
This illustrates why parameter count alone is not a reliable predictor of proving cost. 

To address this, we also report structural statistics of the compiled arithmetic circuits, which are produced automatically by ECC. 
For each circuit, ECC provides the number of addition, multiplication, and constant gates, as well as total number of constraints, among other data. 
These are aggregated into a single measure of circuit complexity called \emph{total cost}, defined as
\begin{multline}
\label{eq:total_cost}
\text{totalCost} \;=\; n_\text{inputs}\cdot C_\text{input} \;+\; n_\text{gates}\cdot C_\text{var} \\ \;+\; n_\text{mul}\cdot C_\text{mul} 
\;+\; n_\text{add}\cdot C_\text{add} \;+\;  n_\text{cst}\cdot C_\text{const},
\end{multline}
where $C_\text{input}, C_\text{var}, C_\text{mul}, C_\text{add}, C_\text{const}$ are fixed configuration weights set by ECC, and $n_\text{inputs}, n_\text{gates}, n_\text{mul}, n_\text{add}, n_\text{cst}$ are the counts of circuit inputs, gates, multiplications, additions, and constants respectively. 
Because each term reflects a distinct kind of gate or variable introduced during circuitization, total cost serves as a compressed proxy for the overall gate inventory.
See Table~\ref{tab:depth_struct}.

\begin{table}[ht]
  \centering
  \resizebox{\columnwidth}{!}{%
\begin{tabular}{rrrrrrr}
\toprule
d & parameters & total cost & constraints & const & add & mult \\
\midrule
1 & 1,607,642 & 9,334,509,820 & 4,335,006 & 163,232 & 16,399,888 & 4,260,596 \\
2 & 405,738 & 9,356,600,728 & 5,416,926 & 200,864 & 20,182,212 & 5,323,700 \\
3 & 408,058 & 9,360,943,969 & 5,583,134 & 204,000 & 21,093,703 & 5,483,636 \\
4 & 410,378 & 9,366,934,139 & 5,749,342 & 207,136 & 22,008,037 & 5,643,572 \\
5 & 412,698 & 9,371,286,878 & 5,915,550 & 210,272 & 22,922,694 & 5,803,508 \\
6 & 415,018 & 9,375,639,182 & 6,081,758 & 213,408 & 23,837,206 & 5,963,444 \\
7 & 417,338 & 9,379,990,879 & 6,247,966 & 216,544 & 24,751,519 & 6,123,379 \\
8 & 419,658 & 9,384,342,586 & 6,414,174 & 219,680 & 25,665,832 & 6,283,315 \\
9 & 421,978 & 9,405,079,425 & 6,580,382 & 222,816 & 26,580,519 & 6,443,252 \\
10 & 424,298 & 9,409,432,188 & 6,746,590 & 225,952 & 27,495,184 & 6,603,188 \\
11 & 426,618 & 9,413,783,895 & 6,912,798 & 229,088 & 28,409,497 & 6,763,124 \\
12 & 428,938 & 9,418,136,193 & 7,079,006 & 232,224 & 29,324,007 & 6,923,060 \\
13 & 431,258 & 9,422,487,918 & 7,245,214 & 235,360 & 30,238,326 & 7,082,996 \\
14 & 433,578 & 9,426,840,126 & 7,411,422 & 238,496 & 31,152,806 & 7,242,932 \\
15 & 435,898 & 9,431,191,251 & 7,577,630 & 241,632 & 32,066,925 & 7,402,868 \\
16 & 438,218 & 9,435,543,474 & 7,743,838 & 244,768 & 32,981,410 & 7,562,804 \\
\bottomrule
\end{tabular}
  }
  \caption{Circuit summary for depth sweep. Columns show depth $d$, parameter count, and ECC circuit data.}
  \label{tab:depth_struct}
\end{table}

\begin{figure}[ht]
  \centering
  \begin{subfigure}[ht]{\columnwidth}
    \centering
    \includegraphics[width=\columnwidth]{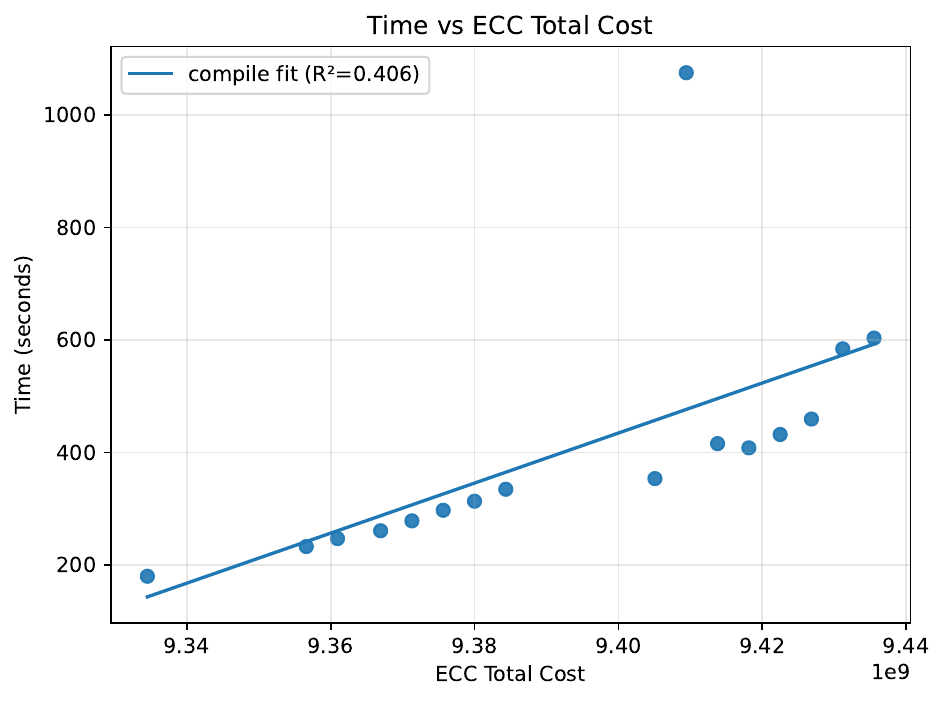}
    \caption{Compile}
  \end{subfigure}

  \medskip

  \begin{subfigure}[ht]{\columnwidth}
    \centering
    \includegraphics[width=\columnwidth]{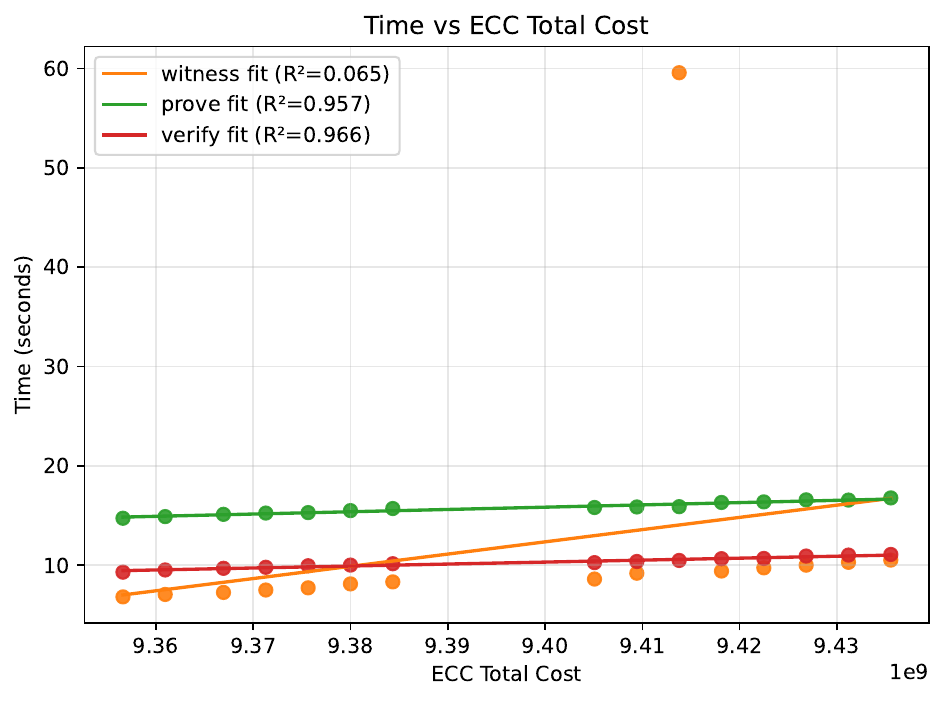}
    \caption{Witness / Prove / Verify}
  \end{subfigure}

  \caption{Depth sweep: runtime vs.~ECC total cost. The bottom panel aggregates non-compile phases.}
  \label{fig:depth_time_vs_cost}
\end{figure}

\begin{figure}[ht]
  \centering
  \begin{subfigure}[ht]{\columnwidth}
    \centering
    \includegraphics[width=\columnwidth]{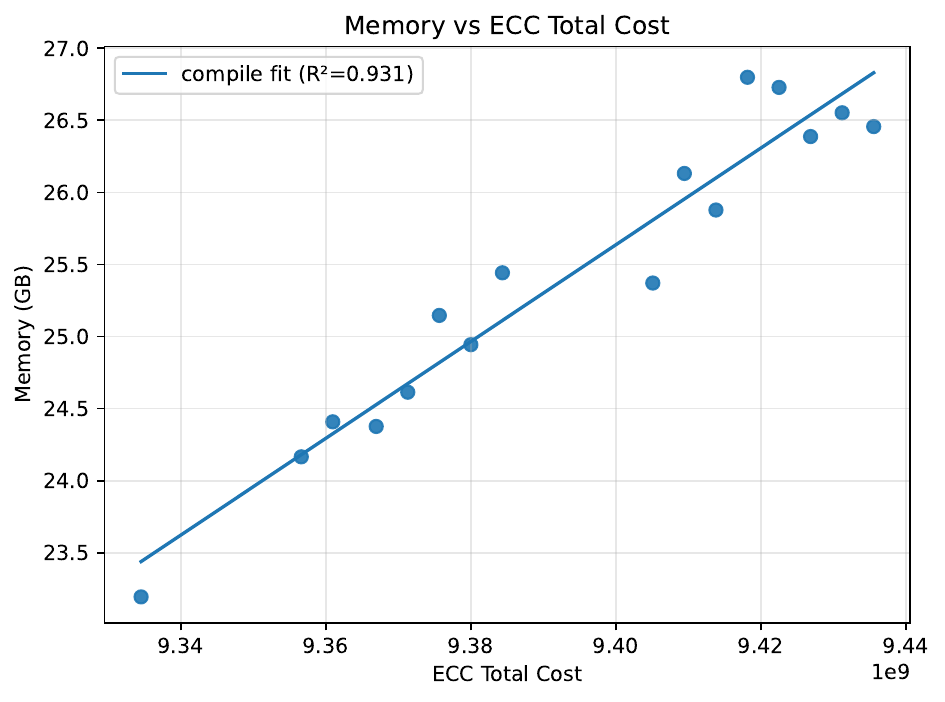}
    \caption{Compile}
  \end{subfigure}

  \medskip

  \begin{subfigure}[ht]{\columnwidth}
    \centering
    \includegraphics[width=\columnwidth]{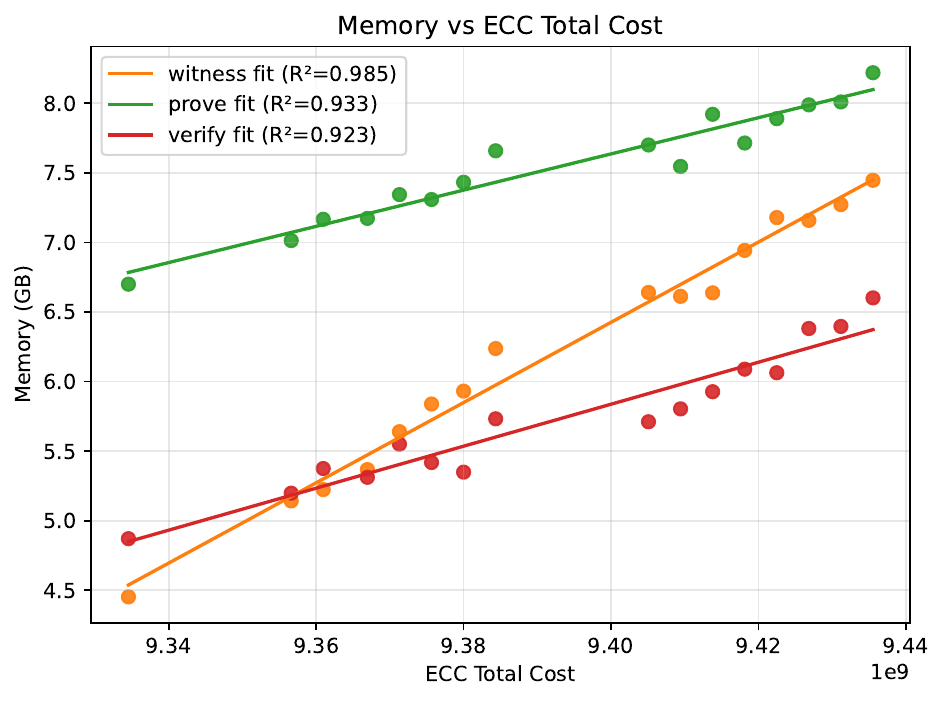}
    \caption{Witness / Prove / Verify}
  \end{subfigure}

  \caption{Depth sweep: peak memory vs.~ECC total cost. The bottom panel aggregates non-compile phases.}
  \label{fig:depth_mem_vs_cost}
\end{figure}

Empirically, total cost proves to be a more reliable predictor of resource usage than parameter count. 
In our depth sweep experiment, for each phase (circuit compilation, witness generation, proving, and verifying), both time and memory scale in an essentially linear fashion with total cost, with $R^2$ values exceeding $0.8$ in simple linear regressions (disregarding some anomalies---see below). 
This suggests that ECC's cost metric effectively captures the true computational complexity of the generated circuits, and provides a meaningful way to extrapolate proving resource requirements beyond the models tested here. 
See Figures~\ref{fig:depth_time_vs_cost} and \ref{fig:depth_mem_vs_cost} for plots. 

\begin{figure}[ht]
  \centering
  \begin{subfigure}[ht]{\columnwidth}
    \centering
    \includegraphics[width=\columnwidth]{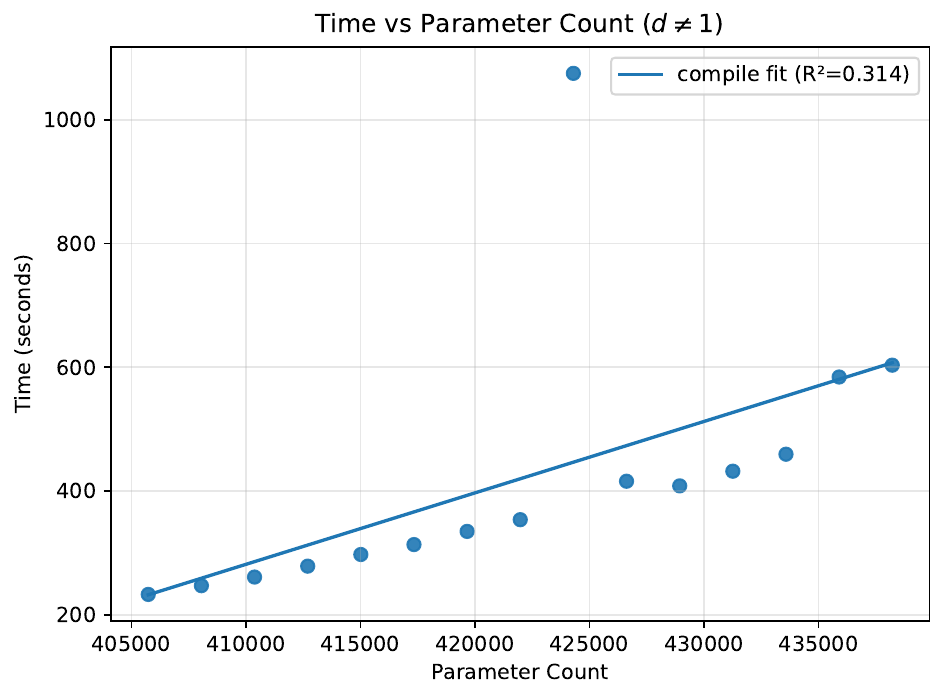}
    \caption{Compile (depth 1 excluded)}
  \end{subfigure}

  \medskip

  \begin{subfigure}[ht]{\columnwidth}
    \centering
    \includegraphics[width=\columnwidth]{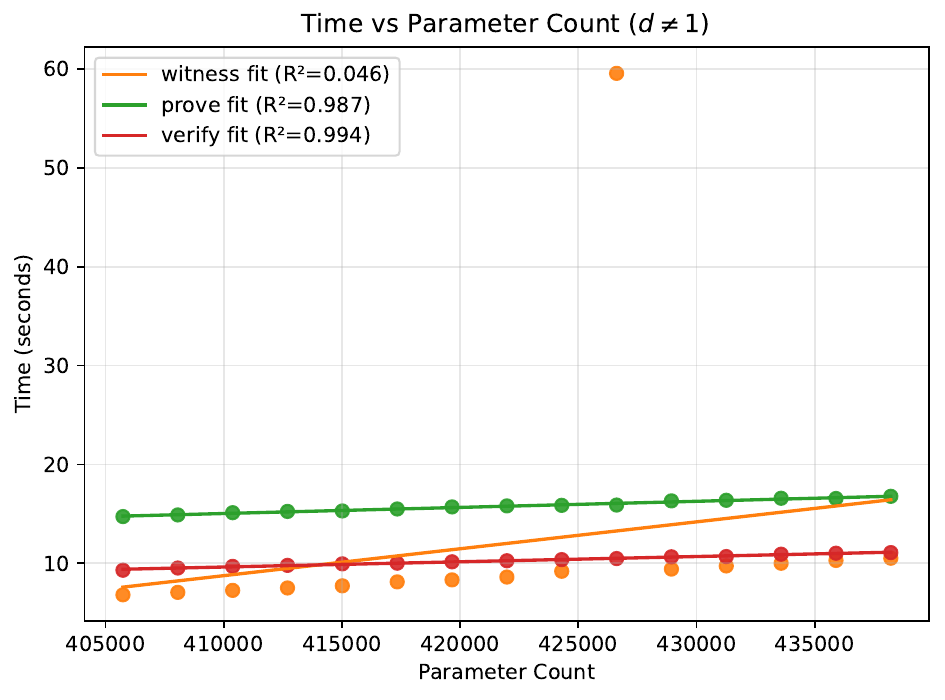}
    \caption{Witness / Prove / Verify (depth 1 excluded)}
  \end{subfigure}

  \caption{Depth sweep: runtime vs.~parameter count. The bottom panel aggregates non-compile phases.}
  \label{fig:depth_time_vs_param}
\end{figure}

\begin{figure}[ht]
  \centering
  \begin{subfigure}[ht]{\columnwidth}
    \centering
    \includegraphics[width=\columnwidth]{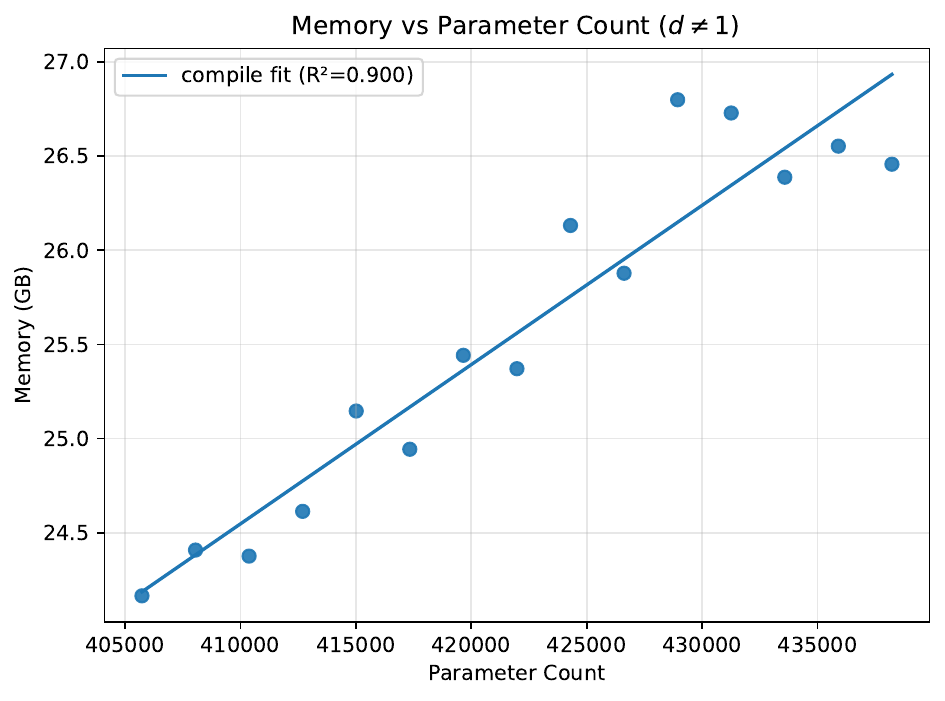}
    \caption{Compile (depth 1 excluded)}
  \end{subfigure}

  \medskip

  \begin{subfigure}[ht]{\columnwidth}
    \centering
    \includegraphics[width=\columnwidth]{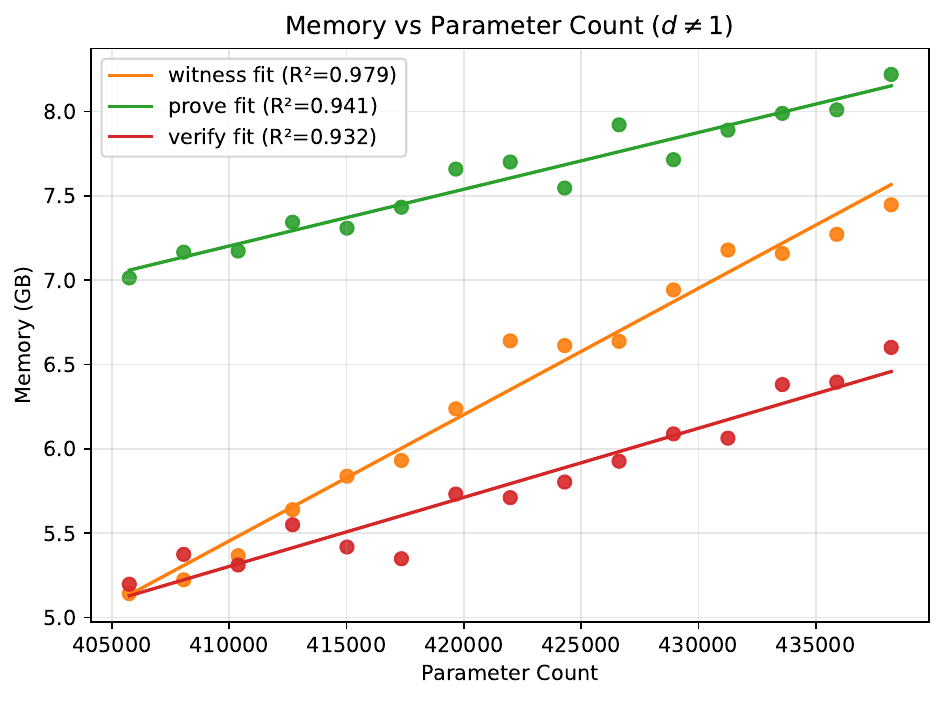}
    \caption{Witness / Prove / Verify (depth 1 excluded)}
  \end{subfigure}

  \caption{Depth sweep: peak memory vs.~parameter count. The bottom panel aggregates non-compile phases.}
  \label{fig:depth_mem_vs_param}
\end{figure}

If we exclude the depth-1 model, whose parameter count is an outlier because the fully connected (FC) tail is fed by a much larger spatial map (only one max-pool) and therefore contains far more weights, we observe a strongly linear relationship between parameter count and both time and memory across depths 2 through 16. 
Intuitively, once the early pooling has regularized the spatial resolution, increasing depth adds blocks of the same type whose parameters and resulting circuit fragments are approximately constant per layer; the total parameter count (and its induced gate count) then grows roughly linearly with depth. 
See Figures~\ref{fig:depth_time_vs_param} and \ref{fig:depth_mem_vs_param} for the corresponding fits.




\begin{figure}[ht]
  \centering
  \begin{subfigure}[ht]{\columnwidth}
    \centering
    \includegraphics[width=\columnwidth]{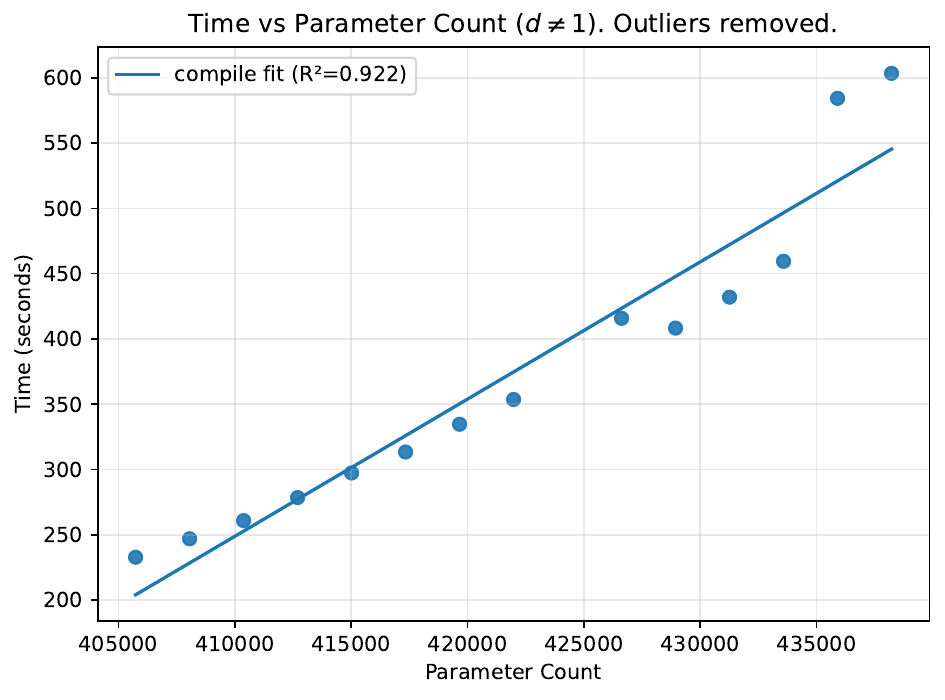}
    \caption{Compile (depth 1 excluded). Outliers removed.}
  \end{subfigure}

  \medskip

  \begin{subfigure}[ht]{\columnwidth}
    \centering
    \includegraphics[width=\columnwidth]{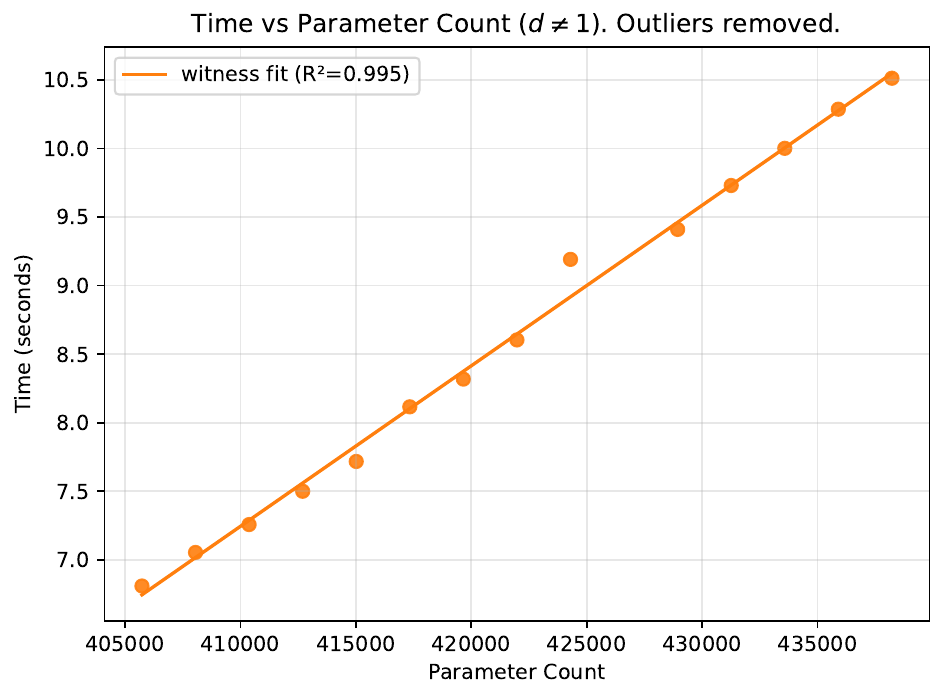}
    \caption{Witness (depth 1 excluded). Outliers removed.}
  \end{subfigure}

  \caption{Depth sweep: runtime vs.~parameter count with outliers removed. (No outliers for prove and verify phases.)}
  \label{fig:depth_time_vs_param_no_outliers}
\end{figure}

\begin{figure}[ht]
  \centering
  \begin{subfigure}[ht]{\columnwidth}
    \centering
    \includegraphics[width=\columnwidth]{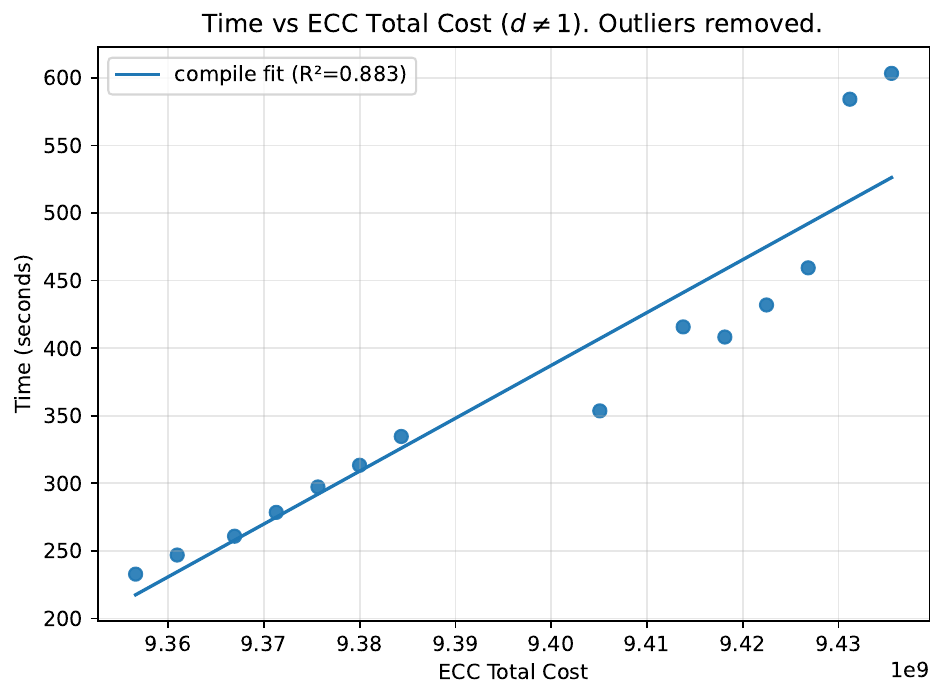}
    \caption{Compile (depth 1 excluded). Outliers removed.}
  \end{subfigure}

  \medskip

  \begin{subfigure}[ht]{\columnwidth}
    \centering
    \includegraphics[width=\columnwidth]{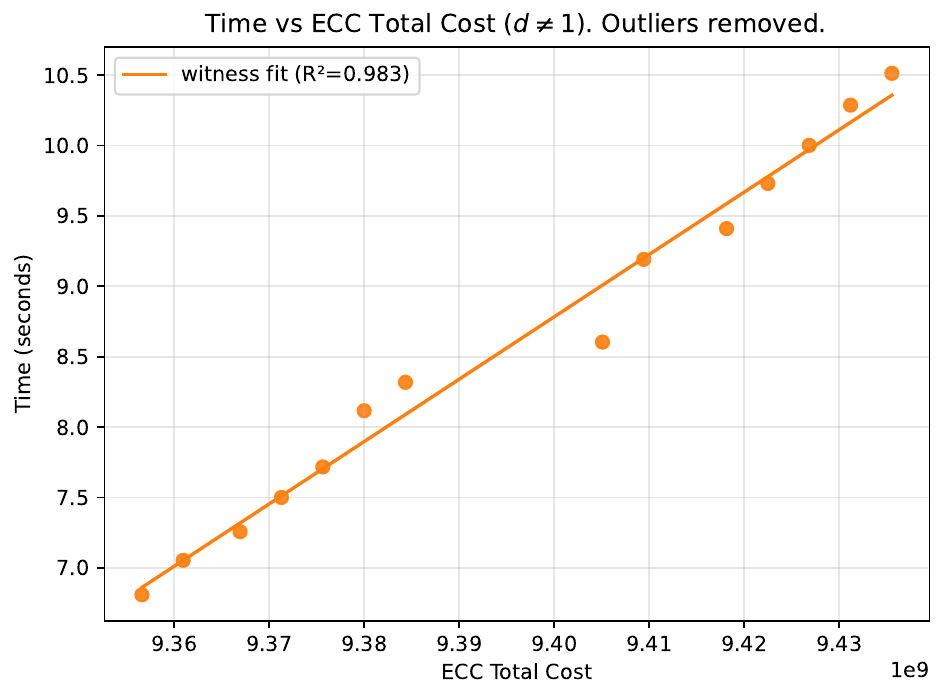}
    \caption{Witness (depth 1 excluded). Outliers removed.}
  \end{subfigure}

  \caption{Depth sweep: runtime vs.~ECC Total Cost with outliers removed. (No outliers for prove and verify phases.)}
  \label{fig:depth_time_vs_cost_no_outliers}
\end{figure}

Two anomalies deserve mention: the depth-10 model showed unusually high compile time ($\sim$1075\,s, versus $\sim$353\,s at depth 9 and $\sim$415\,s at depth 11), while the depth-11 model showed elevated witness generation time ($\sim$59\,s, compared to $\sim$9\,s for neighboring depths). 
We suspect these outliers reflect transient system factors (e.g., resource contention or thermal throttling) rather than circuit complexity itself. 
To identify such anomalies, we applied the Median Absolute Deviation (MAD) method, a robust alternative to standard deviation for measuring spread. 
For each phase, we computed a modified $z$-score
\[
z = 0.6745 \times \frac{|x - \operatorname{median}|}{\operatorname{MAD}},
\]
and flagged any point with $z > 3.5$ as an outlier, following a standard conservative threshold. 
In this dataset, the MAD method flagged only the two anomalies already noted (depth 10 for compile, depth 11 for witness), but it provides a principled, general approach for detecting spurious results in future experiments. 
Because these points skew regression fits, we report both regressions with the outliers included and with them excluded; the latter more accurately reflects the underlying scaling trend. 
See Figure~\ref{fig:depth_time_vs_param_no_outliers} for parameter count results with outliers removed, and Figure~\ref{fig:depth_time_vs_cost_no_outliers} for ECC total cost results with outliers removed.

\subsection{Breadth Sweep}
\label{sec:breadth_sweep}

In our ``breadth sweep'' we fixed a LeNet-style architecture and varied only the input spatial size. 
Concretely, the network used five convolutional blocks with two early max-pool layers, followed by a reshape and a one-layer classifier tail (so $d = 5$, $c = 5$, $p = 2$, $f = 1$); with our ReLU policy this yields one ReLU after each conv and the FC ($r = 6$). 
We evaluated inputs with $h = w \in \{28, 56, 84, 112\}$.
For each model, we ran three complete end-to-end iterations and report the per-phase mean of runtime and peak memory.

Even with fixed kernels and channels, enlarging the spatial tensors increases the amount of convolutional work across multiple layers, and each unit of numeric work typically expands into multiple circuit operations (e.g., scale/quantize handling, range checks, and wiring overhead), not just a single multiplication gate. 
As a result, overall circuit complexity grows super-linearly with input side length, often worse than quadratic in practice, though compiler-level optimizations in Expander can partially mitigate this growth. 
This helps explain why the breadth sweep shows much steeper growth in compile time and memory: at $h = w = 112$ ($3{,}223{,}962$ parameters) the compile phase ran out of memory and failed on our 36 GB machine, whereas all smaller inputs ($28$, $56$, $84$) compiled successfully.
See Table \ref{tab:breadth_time_mem}.


\subsection{Artifact Sizes}
\label{subsec:artifact_sizes}

We also report the sizes of on-disk artifacts (compiled circuit, witness, and proof).
Witness sizes exhibit clear plateaus because the prover allocates workspaces in discrete evaluation domains; models that fall into the same domain serialize to identical witness lengths, so witness size changes in steps rather than smoothly.
Proof sizes grow only weakly with circuit size and therefore vary little across neighboring models.
In contrast, the compiled circuit file tracks circuit complexity more directly and grows steadily with either depth or input size.
These effects explain the near-constant witness/proof sizes in the depth sweep and the step increases in the breadth sweep; at $h = 112$ ($3{,}223{,}962$ parameters) the compile phase exceeded memory on our 36\,GB machine.
See Table~\ref{tab:artifacts_depth_breadth}.


\subsection{Summary}
\label{subsec:benchmarking_summary}

The central finding of our benchmarking is that ECC's \emph{total cost} serves as a reliable predictor of runtime and memory across phases. 
Within a fixed architectural family, raw parameter count can also act as a  reasonable proxy, but when comparing across architectures it becomes misleading (e.g., depth-1 vs.\ depth-2). 
In contrast, total cost consistently captures circuit complexity regardless of architectural differences, making it a more general axis for forecasting resource usage.

The path to larger models is straightforward: adopt well-known circuit optimizations (e.g., lookup-based range checks and probabilistic checks in 
matmul subroutines) to reduce constraint counts, and streamline data 
handling in our pipeline (fewer copies, tighter serialization, and better 
streaming of intermediates) to curb peak RSS and compile time. 
These improvements are incremental, compatible with our current architecture, and aimed directly at the bottlenecks revealed by the measurements.

On the performance side, enabling GPU proving via Expander/ECC's emerging \emph{zkCUDA} support should provide substantial speedups by exploiting parallelism in GKR-style circuits. 
This complements the circuit and pipeline changes above: GPU acceleration targets throughput, while circuit and systems work drive memory footprint and compile efficiency.

Going forward, we will report per-phase improvements relative to this baseline and continue to regress against \emph{total cost}. 
By tracking speedups and peak-memory reductions for each change in isolation, and preserving visibility into failure modes such as OOM at compile, we can make progress measurable and reproducible.

The current study provides both a solid baseline for JSTprove and a dependable control metric (\emph{total cost}) for scaling. 
While our present circuitization and quantization strategies introduce significant overhead, the underlying GKR-based proving system remains well-suited for memory-efficient proofs, underscoring its viability as a foundation for scalable zkML.

\section{Transparency and Auditability}
\label{sec:transparency}

Zero-knowledge proofs reduce the need for trust in individual systems, but confidence in a zkML framework still depends on the visibility of its design and implementation. 
To this end, JSTprove makes all circuit designs and supporting code fully open source. 
Rather than relying on inaccessible components, we publish detailed \emph{zkML Blueprints} \cite{zkmlblueprints2025}, which contain both formal derivations and working implementations. 
This documentation not only allows external validation but also provides a resource for others to study, adapt, and build upon.

Transparency in JSTprove is not limited to releasing code. 
By publishing circuit blueprints, correctness arguments, and design rationales, the project encourages meaningful community engagement. 
Researchers and developers can audit, test, and extend the system, fostering a cycle of collective improvement. 
In this way, JSTprove aims to combine the cryptographic guarantees of zero-knowledge with the practical assurance that comes from open, auditable engineering practices.

\section{Conclusion}
\label{sec:conclusion}

In this work, we introduced JSTprove, an end-to-end pipeline for verifiable machine learning inference. 
By integrating model import, quantization, circuit construction, witness generation, and proof verification into a unified command-line interface, JSTprove lowers the barrier for practitioners to experiment with zkML. 
Our open-source \emph{zkML Blueprints} \cite{zkmlblueprints2025} document the mathematical foundations of core circuit gadgets, ensuring that the system remains transparent, auditable, and extensible. 
Benchmarking across depth- and breadth-swept CNN families demonstrated both the practicality of the approach and the predictive power of ECC's total cost metric, which provides a reliable control axis for scaling to larger models.

At the same time, JSTprove remains a baseline implementation with clear opportunities for refinement. 
Our quantization strategy, while simple and sound, introduces measurable overhead through uniformly applied scaling factors and quotient–remainder checks. 
More adaptive scaling methods could reduce constraint counts while maintaining model fidelity, helping to balance circuit efficiency against numerical precision. 
For the matrix multiplications that dominate fully connected layers, randomized verification methods such as Freivalds' algorithm \cite{Freivalds1977} point to promising reductions in constraint complexity without compromising soundness. 
And on the systems side, the emerging \emph{zkCUDA} framework in Expander suggests a natural direction for parallelizing proof generation: reorganizing circuits into modular, data-parallel components that map efficiently to GPU-like execution environments.

Together, these directions highlight a broader evolution from prototype to production-scale design; one that prioritizes both usability and performance while preserving the mathematical rigor required for cryptographic verification. 
By combining transparency, reproducibility, and sound engineering, JSTprove aspires not only to be a practical toolkit for today's developers but also to serve as a foundation for advancing verifiable AI in real-world applications.

\begin{table}[ht]
\centering
\textbf{Runtime (s) by phase}
\vspace{2pt}
\resizebox{\columnwidth}{!}{%
\begin{tabular}{rrrrrrrrrrr}
\toprule
h & c & p & f & r & parameters & total cost & compile & witness & prove & verify \\
\midrule
28 & 5 & 2 & 1 & 6 & 213,402 & 2,343,195,929 & 61.33 & 2.06 & 5.20 & 3.84 \\
56 & 5 & 2 & 1 & 6 & 815,514 & 9,372,633,329 & 706.60 & 7.67 & 15.14 & 9.78 \\
84 & 5 & 2 & 1 & 6 & 1,819,034 & 18,777,656,794 & 2657.29 & 16.87 & 26.77 & 15.96 \\
\bottomrule
\end{tabular}
}
\medskip\par
\textbf{Peak memory (GB) by phase}
\vspace{2pt}
\resizebox{\columnwidth}{!}{%
\begin{tabular}{rrrrrrrrrrr}
\toprule
h & c & p & f & r & parameters & total cost & compile & witness & prove & verify \\
\midrule
28 & 5 & 2 & 1 & 6 & 213,402 & 2,343,195,929 & 11.665 & 2.002 & 3.748 & 2.839 \\
56 & 5 & 2 & 1 & 6 & 815,514 & 9,372,633,329 & 25.523 & 5.794 & 7.297 & 5.378 \\
84 & 5 & 2 & 1 & 6 & 1,819,034 & 18,777,656,794 & 27.757 & 12.273 & 13.664 & 10.712 \\
\bottomrule
\end{tabular}
}
\caption{Breadth sweep results: runtime and peak memory. $h$ denotes input spatial size (height = width, pixels). Architecture is fixed at depth $d = 5$ with LeNet-like pooling; $c$, $p$, $f$, $r$ are the counts of convolution, max-pool, fully-connected, and ReLU layers. Memory is measured as peak RSS via \texttt{psutil}. \textit{parameters} is the ONNX parameter count. \textit{total cost} is given by \eqref{eq:total_cost}. }
\label{tab:breadth_time_mem}
\end{table}
\begin{table}[H]
  \centering

  \begin{minipage}{\columnwidth}
    \centering
    \textbf{Depth sweep artifact size (MB) by phase}\vspace{2pt}
    \resizebox{\columnwidth}{!}{%
      \begin{tabular}{rrrrrrr}
        \toprule
        d & parameters & total cost & circuit size & witness size & proof size \\
        \midrule
        1 & 1,607,642 & 9,334,509,820 & 871.77 & 256.38 & 0.218346 \\
        2 & 405,738 & 9,356,600,728 & 1074.76 & 256.38 & 0.218346 \\
        3 & 408,058 & 9,360,943,969 & 1121.10 & 256.38 & 0.218346 \\
        4 & 410,378 & 9,366,934,139 & 1167.57 & 256.38 & 0.218437 \\
        5 & 412,698 & 9,371,286,878 & 1214.06 & 256.38 & 0.218437 \\
        6 & 415,018 & 9,375,639,182 & 1260.53 & 256.38 & 0.218437 \\
        7 & 417,338 & 9,379,990,879 & 1307.01 & 256.38 & 0.218437 \\
        8 & 419,658 & 9,384,342,586 & 1353.48 & 256.38 & 0.218437 \\
        9 & 421,978 & 9,405,079,425 & 1399.96 & 256.38 & 0.218712 \\
        10 & 424,298 & 9,409,432,188 & 1446.45 & 256.38 & 0.218712 \\
        11 & 426,618 & 9,413,783,895 & 1492.92 & 256.38 & 0.218712 \\
        12 & 428,938 & 9,418,136,193 & 1539.40 & 256.38 & 0.218712 \\
        13 & 431,258 & 9,422,487,918 & 1585.87 & 256.38 & 0.218712 \\
        14 & 433,578 & 9,426,840,126 & 1632.35 & 256.38 & 0.218712 \\
        15 & 435,898 & 9,431,191,251 & 1678.81 & 256.38 & 0.218712 \\
        16 & 438,218 & 9,435,543,474 & 1725.29 & 256.38 & 0.218712 \\
        \bottomrule
      \end{tabular}
    }
  \end{minipage}

  \vspace{0.75em}

  \begin{minipage}{\columnwidth}
    \centering
    \textbf{Breadth sweep artifact size (MB) by phase}\vspace{2pt}
    \resizebox{\columnwidth}{!}{%
      \begin{tabular}{rrrrrrr}
        \toprule
        h & parameters & total cost & circuit size & witness size & proof size \\
        \midrule
        28 & 213,402 & 2,343,195,929 & 307.56 & 64.09 & 0.115410 \\
        56 & 815,514 & 9,372,633,329 & 1234.04 & 256.38 & 0.218437 \\
        84 & 1,819,034 & 18,777,656,794 & 2782.19 & 512.86 & 0.297295 \\
        112 & 3,223,962 & -- & -- & -- & -- \\
        \bottomrule
      \end{tabular}
    }
  \end{minipage}

  \caption{Artifact sizes (MB) across depth and breadth sweeps. \emph{Top:} depth sweep artifact sizes by depth $d$ for fixed input $1{\times}4{\times}56{\times}56$; witness sizes plateau within domain buckets while circuit size grows with depth. \emph{Bottom:} breadth sweep artifact sizes by input side length $h$. At $h{=}112$, the compile phase exceeded memory, hence ``--''.}
  \label{tab:artifacts_depth_breadth}
\end{table}

\FloatBarrier

\section{Acknowledgments}
\label{sec:acknowledgments}

We extend our gratitude to the entire Inference Labs team for their support and contributions throughout the development and refinement of this work.

\balance
\printbibliography[title=References,heading=bibintoc]

\pagebreak

\onecolumn
\appendix

\section{Code}
\label{sec:code}

\begin{lstlisting}[language=bash, style=boxed, caption={Example JSTprove CLI usage}, label={lst:cli_demo}]
# 1. Compile -> circuit + quantized ONNX
jst compile \
  -m python/models/models_onnx/lenet.onnx \
  -c artifacts/lenet/circuit.txt

# 2. Witness -> reshape/scale inputs, run model, write witness + outputs
jst witness \
  -c artifacts/lenet/circuit.txt \
  -i python/models/inputs/lenet_input.json \
  -o artifacts/lenet/output.json \
  -w artifacts/lenet/witness.bin

# 3. Prove -> witness -> proof
jst prove \
  -c artifacts/lenet/circuit.txt \
  -w artifacts/lenet/witness.bin \
  -p artifacts/lenet/proof.bin

# 4. Verify -> check the proof
jst verify \
  -c artifacts/lenet/circuit.txt \
  -i python/models/inputs/lenet_input.json \
  -o artifacts/lenet/output.json \
  -w artifacts/lenet/witness.bin \
  -p artifacts/lenet/proof.bin
\end{lstlisting}

\begin{remark}
Paths are user-specified; here we show the included LeNet-style demo for concreteness. 
Any supported ONNX model with a matching input JSON can be substituted.
\end{remark}

\begin{lstlisting}[language=Rust, style=boxed, caption={Minimal implementation of a range check via bitstring decomposition, illustrating the use of ECC’s Rust API}
, label={lst:range_check}]
use expander_compiler::frontend::{Config, RootAPI, Variable};

pub fn unconstrained_to_bits<C: Config, Builder: RootAPI<C>>(
    api: &mut Builder,
    input: Variable,
    n_bits: usize,
) -> Vec<Variable> {
    assert!(n_bits > 0, "n_bits must be > 0");
    let mut bits = Vec::with_capacity(n_bits);
    let mut cur = input;
    for _ in 0..n_bits {
        let bit = api.unconstrained_bit_and(cur, 1u32);
        bits.push(bit);
        cur = api.unconstrained_shift_r(cur, 1u32);
    }
    bits
}

pub fn assert_is_bitstring_and_reconstruct<C: Config, Builder: RootAPI<C>>(
    api: &mut Builder,
    bits_le: &[Variable],
) -> Variable {
    let mut acc = api.constant(0u32);
    for (i, &b) in bits_le.iter().enumerate() {
        api.assert_is_bool(b);
        let w = api.constant(1u32 << (i as u32));
        let term = api.mul(w, b);
        acc = api.add(acc, term);
    }
    acc
}

// Range check for x in [-2^{kappa-1}, 2^{kappa-1}-1].
// Shifts x by 2^{kappa-1}, bit-decomposes the result, and enforces reconstruction.
pub fn range_check_signed_kappa_bits<C: Config, Builder: RootAPI<C>>(
    api: &mut Builder,
    x: Variable,
    kappa: usize,
) {
    let shift = api.constant(1u32 << ((kappa - 1) as u32));
    let x_sharp = api.add(x, shift);             // x^{sharp} = x + 2^{kappa-1} (mod p)
    let bits = unconstrained_to_bits(api, x_sharp, kappa);
    let recon = assert_is_bitstring_and_reconstruct(api, &bits);
    api.assert_is_equal(x_sharp, recon);         // enforces 0 <= x^{#} < 2^{kappa}
}
\end{lstlisting}

\begin{remark}
Listing~\ref{lst:range_check} illustrates only the core use of 
ECC's \texttt{RootAPI} for range checking. 
Our production implementation includes additional safeguards such as 
bounds checks (e.g., ensuring $2^{\kappa} \le p$) and overflow handling. 
Here we show the minimal version to make the construction transparent.
\end{remark}

\twocolumn

\end{document}